\documentstyle[11pt,amssymbols,times,psgreek]{article}
\newcommand{\be}{\begin{equation}\abovedisplayskip 5mm
 \belowdisplayskip 5mm\abovedisplayshortskip 5mm
  \belowdisplayshortskip 5mm\jot 3mm}
\newcommand{\ee}{\end{equation}}
\newcommand{\ba}{\begin{array}}

\newcommand{\ea}{\end{array}}
\newcommand{\bea}{\begin{eqnarray}\abovedisplayskip 5mm
 \belowdisplayskip 5mm\abovedisplayshortskip 5mm
  \belowdisplayshortskip 5mm\jot 3mm}
\newcommand{\eea}{\end{eqnarray}}
\newcommand{\lra}{\longrightarrow}
\newcommand{\ra}{\rightarrow}
\newcommand{\la}{\label}

\newcommand{\re}{\ref}
\newcommand{\p}{\partial}
\newcommand{\ol}{\overline}
\newcommand{\ti}{\tilde}
\begin{document}
\begin{titlepage}
November 1991\hfill PAR-LPTHE 91/57
%\begin{flushright} MAINZ-TH 91/yy\end{flushright}
\vskip 3cm\centerline{\bf 4-DIMENSIONAL SPONTANEOUSLY
BROKEN $U(1)_L$ GAUGE THEORIES }
\medskip
\centerline{\bf WITH A COMPOSITE WESS-ZUMINO FIELD}
\medskip
\centerline {(Revised February 1992)}
\vskip 1cm
\centerline{B. Machet}
\vskip 3mm
\centerline{{\em Laboratoire de Physique Th\'eorique et
Hautes Energies}
\footnote{Laboratoire associ\'e au CNRS URA 280.}
\footnote{Postal address: LPTHE tour 16-1er
 \'etage, Universit\'e P. et M. Curie, BP 126, 4 place
 Jussieu,\\ F 75252 PARIS CEDEX 05 (France).}, {\em Paris}}
 %\vskip 6mm
 %\centerline{G. Thompson}
 %\vskip 3mm
 %\centerline{{\em Institut f\"ur Physik, Universit\"at
 %Mainz} \footnote{Postal address: Institut f\"ur
 %Physik, Johannes Gutenberg Universit\"at, Staudinger Weg
 %7, Postfach 39 80, D 6500 MAINZ (FRG).}}
 %
 %
 \vskip 3cm
 {\bf Abstract:}\\
Some spontaneously broken gauge theories with
left couplings to fermions, like the abelian  model
that we propose here, can be endowed with a composite scalar sector
and Wess-Zumino field ; their quantization in the
functionnal integral formalism accordingly requires the
introduction of constraints that, together with the breaking of the
gauge symmetry by the scalars, and  among other consequences,
give the Higgs field and the fermions (quarks) infinite masses;
this makes them unobservable.
Gauge invariance and unitarity are achieved through a
derivative coupling of the W-Z field to the fermionic current;
the anomaly gets cancelled in the above infinite fermion mass
limit.  We show how the problems of renormalizability are evaded at the
one-loop level by resumming diagrams at the ladder approximation
and  reshuffling the perturbative series, and because
the fermionic current is conserved.  The Wess-Zumino field
can be "gauged away" to become the 3rd polarization of the massive gauge
field; the pseudoscalar partner of the Higgs, tightly linked to the W-Z
field, behaves like an abelian pion. In particular, no extra
scale of interaction has to be introduced, unlike in "technicolour"
theories.
Problems concerning the leptonic sector are only mentionned.
\smallskip

 \vskip 1cm
 %\centerline{\rule {3cm} {0.2pt}}
 \vfill
 \centerline{Work supported by CNRS}
 \end{titlepage}

\section {Introduction.}
  One of the keystones of today's particle physics is the
  cancellation of gauge anomalies \cite{ano} as a condition for
 renormalizability and hence for calculability. The
  pioneering works \cite{bim} showed indeed that achieving both
 unitarity and renormalizability was impossible in the
  strict framework of a spontaneously broken gauge theory
  with a standard scalar sector (by "standard" we mean a
  fondamental Higgs and its Goldstone partner(s)). Curing
  this desease required the choice of an appropriate gauge
  group \cite{go} or/and a subtle cancellation between
different
  kinds of fermions \cite{bim}. Hence today's dogma of a strict
 parallel between quarks and leptons to cancel the
  potential $U(1)$ anomaly of the Standard Model \cite{gsw}.

  The quest for gauge invariance despite the presence of
  anomalies was revived in the recent years. It was shown
  that the symmetry could be  restored with the introduction
  of  extra scalar fields and their additional
  contributions to the Lagrangian, involving in particular
  a Wess-Zumino term with a reversed sign. First
  introduced by hands \cite{fasha}, it was then shown
  \cite{bsv} that, in
  the functionnal integral formalism, this appeared as a
  natural consequence of the integration over the whole
  gauge group and not only over the orbit space. However,
  though gauge invariance was indeed recovered, it was soon
  recognised \cite{bmv} that the original problem of achieving
both
  unitarity and renormalizability was unsolved. The source
  of the difficulty lies in that this degree of
  freedom (in the $U(1)$ case) appears in fact as a
  dimensionless scalar field.
  Power counting renormalizability requires its propagator
  to behave as $1/k^4$, which unavoidably reintroduces in
 the theory an uncancelled ghost pole.

  So, though no definitive proof has been established yet,
  there is a widespread belief that the problem cannot be
  solved strictly along these lines.

The step that we take here is to abandon the idea that the
extra scalar field $\xi$ (called hereafter the Wess-Zumino
\cite{wz}
field) is an independent degree of freedom; when the theory
 is spontaneously broken by scalar fields
  $(H,\phi)$ such that $\langle H\rangle = v \not= 0$, and
  when the gauge generator and the unit matrix form an
  associative algebra, there is a unique solution $\xi
=\xi(H,\phi)$ such that
  $\xi$ transforms non-linearly by a gauge transformation
  on the fermion (quark) fields if  $H$ and $\phi$ are
themselves taken as composite fermion operators.

This composite Wess-Zumino (W-Z) field is introduced
through a derivative coupling to the fermionic
current, making the Lagrangian a function of the generic gauge
 field $A_\mu = \sigma_\mu -(1/g)\ {\p_\mu\xi/v}$ only.
This ensures gauge invariance and unitarity; the latter can be
proved either by showing the cancellation of the residue of the
ghost pole in the gauge field propagator (that we shall do here
at the tree level), or by going to the equivalent of the
"unitary gauge", as usually done in the Standard Model
\cite{ab-lee}.

Integrating over non independent degrees of freedom (like $H$,
$\phi$) requires introducing the appropriate constraints in the
functionnal integral formalism of quantization. They are shown
to affect the model in 2 respects:\\
\quad - once they are exponentiated, and together with the fact
that the gauge symmetry is
"spontaneously broken" by $\langle H\rangle =v$, they limit the
number of asymptotic states by giving
the quarks and the Higgs field infinite masses, making them
unobservable;\\
\quad - they provide an important step towards renormalizability; the
above limit of infinite fermion mass yields the
cancellation of the anomaly and the conservation of the fermionic
current; furthermore, we show at the one-loop level that the the 4-fermi
couplings arising from the exponentiation of the constraints become
harmless when the perturbative expansion is reshuffled and infinite series
of "ladder" diagrams resummed.

Another immediate consequence of choosing the scalars as
composite is to unify the usually disconnected 2 phenomena of
symmetry breaking:\\
\quad - gauge symmetry breaking, triggered by  $\langle
H\rangle = v \not= 0$;\\
\quad - chiral symmetry breaking \cite{csb}, triggered by fermion
"condensation"  $\langle \overline\Psi \Psi \rangle
  \not=0$.

Our analysis of the lepton sector will stay incomplete; we
couple the leptonic fermionic current in the same way as the
hadronic current and only stress the consequences concerning
the decays of $\phi$, the pseudoscalar partner of the Higgs; we
show that $\phi$ behaves like an abelian pion whose leptonic
decays come out as usually computed with the so-called "PCAC"
hypothesis \cite{csb}, without having to introduce, like
 in "technicolour" theories \cite{sw},
 an extra scale of interaction; the (anomalous) decays
into 2 photons are now mediated by leptonic loops (due to the
above mentionned cancellation of the anomaly in the quark
sector).\\
The leptonic sector introduced in that way stays
non-renormalizable. We leave a more detailed study
of this question to a forthcoming work \cite{bm1}.

\section {The Wess-Zumino field as a fermionic bound state.}

Let us consider a spontaneously broken $U(1)_L$ gauge
theory. The explicit Lagrangian will appear in the next
section.\\
The gauge group $U(1)_L$ acts on the fermions and  on
the gauge field
$\sigma_\mu$, coupled by a $V-A$ law. There are N
fermions (quarks) with the same coupling,
N is the number of ``flavours''.
We describe them by an N-vector $\Psi$
on which acts the $U(N)_L\times U(N)_R$ chiral group. We
embed $U(1)_L$ into $U(N)_L\times U(N)_R$ and the generator
\be {\Bbb T}_L={1-\gamma_5\over 2}{\Bbb T}\ee
is an $N\times N$ matrix.\\
$Tr \{{\Bbb T},{\Bbb T}\}{\Bbb T}$ can be non-vanishing,
and so the fermionic current
\be J_\mu^\psi = \overline\Psi\gamma_\mu {\Bbb T}_L\Psi =
{1\over 2}(V_\mu -  J _\mu ^5)\ee
can be anomalous.\\
An important condition for the following is
\be {\Bbb T}^2= 1,\la{eq:assoc}\ee
which is the simplest case when a gauge generator (${\Bbb
T}$)
and the unit matrix form an associative algebra (the
generalization to the non-abelian case will be dealt
with in \cite{bm2}).\\
We shall choose in the following ${\Bbb T}$ as the unit
matrix
\be {\Bbb T}= 1_N,\ee
but other cases can be considered as well.\\
When (\re{eq:assoc}) is satisfied, there exists a particular
two-dimensional scalar representation of the gauge group :
\be \Phi=(H,\phi)={v\over\mu^3}(\overline\Psi  1
\Psi,\Psi\gamma_5 {\Bbb T}\Psi).\ee
(H is real, $\phi$ purely imaginary and both have
dimension 1. We shall also use $\varphi$ such that $\phi
= i\varphi)$\\
Indeed, the action of $U(1)_L$ on $\Phi$ is deduced from
that on $\Psi $ and we have, using (\re{eq:assoc}):
\be\begin{array}{c} {\Bbb T}_L.\phi =-H,\\
     {\Bbb T}_L.H =-\phi.\end{array}\la{eq:gract}\ee
The gauge theory is spontaneously broken by
\be \langle H\rangle=v,\ee
which is equivalent to
\be\langle\overline\Psi\Psi\rangle=\mu^3.\ee
It is an immediate consequence of our construction that
the so-called
``gauge breaking'' and ``chiral breaking'' are the same
phenomenon, the mechanism of which we will not investigate
here.

Writing
\be H= v+h,
\la{eq:shift1}\ee
let us define $\tilde H $ and $\xi$, both real, by :
\be H+i\varphi =e^{i\frac{\xi}{v}{\Bbb T}_L}.\
\tilde H,\la{eq:chvar1}\ee
with
\be \tilde H = v+\eta.\la{eq:shift2}\ee
The solution of (\re{eq:chvar1}) is
\be\left\{\ba{lcl}
H\sin{\xi\over v}+\varphi\cos{\xi\over v} &=& 0, \\
\tilde H &=& H\cos{\xi\over v}-\varphi\sin{\xi\over v}.
\ea\right .\la{eq:chvar2}\ee
$\eta$ and $\xi$ can be expressed uniquely as series in
$h/v$ and $\varphi/v$:
\be\ba{lcl}
\xi &=& -\varphi(1-{h\over v})+\cdots,\\
\eta &=& h+{\varphi^2\over 2v}+\cdots .
\ea\ee
 The laws of transformation of $\tilde H$ and $\xi$ can be
 deduced from (\re{eq:gract}) and (\re{eq:chvar2}):\\
\be
when\qquad \Psi \lra e^{-i\theta {\Bbb T}_L} \Psi,
\ee
\be\left\{\ba{lcl}
\xi &\lra& \xi -\theta v,\\
\tilde{H} & & invariant.
\ea\right .\la{eq:trans}\ee
A gauge transformation induces a translation on the field
$\xi$ and (\re{eq:trans}) corresponds to a non-linear realization
of the gauge symmetry:
\be e^{i{\xi\over v}{\Bbb T}_L} \lra e^ {-i\theta {\Bbb
T}_L}\  e^ {i{\xi\over v}{\Bbb T}_L}.\ee
$\xi$ consequently appears as a natural candidate for a
Wess-Zumino field.
Furthermore, as in a gauge transformation, the variation of
$\sigma _\mu$ gets cancelled by that of $-(1/
g)\ \p_\mu \xi/ v$,  the generic (gauge invariant)
gauge field
\be A_\mu = \sigma _\mu -{1\over g}{\p _\mu\xi\over v}
\la{eq:amu}\ee
can be used to quantize the theory along the lines of \cite{bsv}.

\section{Quantizing.}
  \subsection{Gauge symmetry.}

Let the Lagrangian ${\cal L} (x)$ be
  \be\ba{ccl} {\cal L} (x)&=& -{1\over 4} F_{\mu \nu}F^{\mu\nu}
 +i\overline\Psi \gamma^\mu(\p_\mu -ig(\sigma _\mu
  -(1/ g)\ \p_\mu \xi/v){\Bbb T}_L)\Psi\\ & &
  +{ 1\over 2} \p_\mu \tilde H \p^\mu \tilde H +{1\over 2}
  g^2(\sigma _\mu -(1/g)\ \p_\mu \xi/v)^2 \tilde
  H^2 - V(\tilde H^2). \ea\la{eq:L}\ee
$V(\tilde H^2)$ is the scalar potential, a polynomial of
degree 4 in $\tilde H$.

Though it is not intended to be used for computations, we
introduced at the beginning this form of the Lagrangian to
stress a first symmetry which appears at this (classical) level:
as ${\cal L}$ is a function of $A_\mu$ (\re{eq:amu})
only, it is invariant by the transformation
  \be
  \left\{\ba{lcl}
  \xi &\lra & \xi -\theta v,\\
  \sigma_\mu &\lra & \sigma_\mu -(1/g)\  \p_\mu \theta.
  \ea\right .
  \la{eq:clasym}\ee
This invariance is however not that of the theory that we shall
describe because of
the constraints linking the scalars to the fermions: a
transformation on the latter is also needed.

We shall describe the full (quantum) theory in terms of the
variables $\phi$ and $H$, instead of $\xi$ and $\tilde H$,
introduce the 2 constraints linking $\phi$ and $H$ to the
fermions, and functionnally integrate
  \be Z= \int {\cal D} \Psi{ \cal D} \overline\Psi{\cal D}H
  {\cal D} \phi {\cal D} \sigma_\mu \  e^{i\int d^4 x {\cal L}(x)}
 \prod _x \delta(C_H(x))\prod _x \delta (C_\phi(x)),\la{eq:Z} \ee
where
\be
C_H= H-{v\over \mu^3}\ol\Psi\Psi,
\la{eq:CH}\ee
\be
C_\phi=\phi -{v\over \mu^3}\ol\Psi\gamma_5{\Bbb T}\Psi.
\la{eq:Cphi}\ee
The product\  $\prod _x \delta(C_H(x))\prod _x \delta (C_\phi(x))$\
 is not invariant by (\re{eq:clasym}).

$\ti H$ and $\xi$ are constructed such that
\be
\ti H^2 = H^2 -\phi^2,
\ee
and
\be
+{1\over 2} \p_\mu \tilde H \p^\mu \tilde H +{1\over
2}g^2(\sigma_\mu -{1\over g}\p_\mu{\xi\over v})^2\tilde
H^2 = {1\over 2}(D_\mu H D^\mu H -D_\mu\phi D^\mu\phi),
\ee
with
\be\ba{lclcl}
D_\mu H &=& \p_\mu H -ig\sigma_\mu {\Bbb T}_L.H&=& \p_\mu H
+ig\sigma_\mu\phi \\
D_\mu\phi &=& \p_\mu\phi -ig\sigma_\mu {\Bbb T}_L.\phi &=&
\p_\mu\phi +ig\sigma_\mu H,
\ea\ee
such that ${\cal L}(x)$ can also be written
\be\ba{lcl}
{\cal L} &=&
-{1\over 4}F_{\mu\nu}F^{\mu\nu} +i\overline\Psi
\gamma^\mu(\p_\mu -ig\sigma_\mu {\Bbb T}_L)\Psi\\
& &
+{1\over 2}\left((D_\mu H)^2 -(D_\mu \phi)^2\right) -V(H^2
-\phi^2)\\
& & -(\p_\mu\xi/ v)\  J^\mu_\psi;
\ea\la{eq:Lsm}\ee
we recognize in (\re{eq:Lsm}) the Lagrangian of an
abelian "standard model" to which has been added the coupling
(after integrating by parts)
\be
(\xi/ v)\ \p_\mu J^\mu_\psi,
\la{eq:WZ}\ee
which will play the role of a Wess-Zumino term with a reversed
sign.\\
The theory defined by (\re{eq:Z}) and (\re{eq:Lsm}) has now the
(quantum) invariance:
  \be
  \left\{\ba{lcl}
  \Psi &\lra & e^{-i\theta {\Bbb T}_L}\Psi,\\
  \sigma_\mu &\lra & \sigma_\mu -(1/g)\  \p_\mu \theta,\\
  \xi & \lra & \xi - \theta v,
  \ea\right .
  \la{eq:qsym}\ee
because  it leaves the product $\delta(C_H)\  \delta(C_\phi)$
invariant, and the non-invariance of the action due to
(\re{eq:WZ}) is compensated by
the non-invariance of the fermionic measure \cite{fuji}: one recovers
 the Ward Identities of an anomaly-free naive standard model without a
W-Z field. We refer the reader to Appendix A for more details.\\
As a consequence, we shall eventually fix the gauge by adding
\be
 {\cal L}_{gf} = -{1\over{2\alpha}}(\p_\mu\sigma^\mu)^2.
  \ee
to the Lagrangian. This can be done with the standard
Faddeev-Popov procedure \cite{fadpop}. As it involves now the full
transformation (\re{eq:qsym}), we repeat it for completeness
in the Appendix B, which shows in particular that the integration
over the gauge group keeps factorizing. As the ghost decouples,
we shall not comment in the following on the associated
Faddeev-Popov lagrangian ${\cal L}_{FP}$.

  \subsection{Unitarity.}

We show in two ways below how unitarity occurs:\\
\quad - the first goes along the lines of \cite{bim} and deals
with diagrams at the tree level;\\
\quad - the second shows how to switch, by a change of variables in
the functionnal integral, to a form of the theory where the W-Z
field is "gauged away" and the massive gauge field propagates like
in the unitary gauge.

We start by looking at the unitarity problem as was exposed in
\cite{bim}, in the Landau gauge, and show that, in the theory
defined with ${\cal L} +{\cal L}_{gf}+{\cal L}_{FP}$,
 no violation of unitarity occurs at the tree level. The Landau
gauge is specially simple since no $\sigma_\mu\p^\mu\xi$
coupling occurs.\\
When $\tilde H$ gets a vacuum expectation value (\re{eq:shift2}),
 $\sigma_\mu$ becomes massive with mass
 \be
 M=gv,
 \la{eq:gmass}\ee
 and a kinetic term for $\xi$ is produced:
 \be
 {1\over 2}\p_\mu\xi\ \p^\mu\xi.
 \ee
 $\xi$ is a Goldstone particle. The gauge field propagator is
 \be
 D_{\mu\nu}^\sigma = -i\left(\frac{g_{\mu\nu}-k_\mu
 k_\nu/k^2}{k^2 -M^2}\right),
 \ee
 which has the usual ghost pole at $k^2=0$. Since
 $\sigma_\mu$ is coupled to a potentially non-conserved
 current, this
 can break unitarity, unless the corresponding residue is
 cancelled by that of another massless particle. This is
 precisely the role that $\xi$ plays here, as shown on
 fig.(1).
\newpage
\null
 \vskip 5truecm
 {\centerline{\em Fig.1: Cancellation of the ghost residue
 by the Goldstone in the Landau gauge.}}
 \vskip 5mm

 In other gauges $(\alpha\not= 0)$, there exists a
 non-diagonal coupling
 \be
 -M \sigma_\mu\p^\mu\xi.
\ee
 The resummed gauge field propagator (see fig. (2)) is
 \be
 \tilde D_{\mu\nu}^\sigma = -i\left(\frac{g_{\mu\nu}-k_\mu
 k_\nu/k^2}{k^2 -M^2}+\alpha \frac{k_\mu
 k_\nu}{k^4}\right),
 \ee
 and the resummed $\xi$ propagator (see fig.(2))
 \be
 \tilde D^\xi = i\ \frac{k^2 -\alpha M^2}{k^4}.
 \ee
 %
%\newpage\null
 \vskip 6.5 truecm
 {\centerline{\em Fig.2: Resumming the $\sigma_\mu$ and
 $\xi$ propagators.}}
 \vskip 5mm
\noindent In addition to the diagrams of fig.(1), we have now to
 look also at those of fig.(3)  and we check that the
 cancellation of the residue of the ghost pole again
 occurs.
\newpage
\null
 \vskip 5truecm
 {\centerline{\em Fig.3: Other diagrams
 for ghost cancellation  in a general $\alpha$-gauge.}}
 \vskip 5mm
 We thus conclude that unitarity is achieved at tree level
 despite the presence of an anomaly when the Goldstone
 $\xi$ is introduced according to our procedure.
\medskip

We make now a more complete analysis showing how one can go to
the equivalent of the unitary gauge \cite{ab-lee} in the
 Standard Model.\\
In Z defined by (\re{eq:Z}), we transform by exponentiation the
2 constraints into
 \be\ba{ccl}
 {\cal L}_c& =&
\lim_{\beta\ra 0}\frac{-\Lambda^2}{2\beta}
\left(H^2 -\phi^2 -{2v\over \mu^3}(H\overline\Psi\Psi-\phi
\overline\Psi\gamma_5 {\Bbb T} \Psi)+
 {v^2\over\mu^6}\left((\overline\Psi\Psi)^2
-(\overline\Psi\gamma_5 {\Bbb T} \Psi)^2\right)\right)\\
 & =&
\lim_{\beta\ra 0}\frac{-\Lambda^2}{2\beta}
  \left(\tilde H^2-{2v\over \mu^3}\tilde
  H(\overline\Psi\Psi \cos{\xi\over v}
  +i\overline\Psi\gamma_5 {\Bbb T} \Psi \sin{\xi\over
  v})+{v^2\over\mu^6}\left((\overline\Psi\Psi)^2
  -(\overline\Psi\gamma_5 {\Bbb T} \Psi)^2\right)\right),
\ea\la{eq:Lc}\ee
where $\Lambda$ is an arbitrary mass scale,
and go to the variables $\xi$ and $\ti H$.\\
$Z$ can now be written
\be
Z=\int {\cal D} \Psi{ \cal D} \overline\Psi
{\cal D} \tilde H {\cal D} \xi {\cal D} \sigma_\mu
\ J_1\    e^{i\int d^4 x {\cal L}(x)+{\cal L}_c(x)},
\la{eq:Z2}\ee
where $J_1$ is the Jacobian coming from the change of variables
\be
{\cal D} H {\cal D} \phi = J_1\  {\cal D} \tilde H {\cal D}
\xi,
\ee
which can be expressed as
\be
J_1 =\prod _x {\ti H\over v}=
e^{\delta^4(0)\int d^4x\sum_n (\eta/v)^n}.
\la{eq:J1}\ee
We finally perform the changes of variables
\be
  \left\{\ba{lcl}
  \Psi &\lra & e^{-i(\xi/v){ \Bbb T}_L}\Psi,\\
  H+i\varphi &\lra & e^{-i(\xi/v) {\Bbb T}_L}.\ (H+i\varphi);
  \ea\right .
\la{eq:chvar}\ee
the last line in eq.(40) is identical to the transformation (3.7a) of
Abers and Lee \cite{ab-lee}, and is equivalent in terms of
$\ti H$ and $\xi$ to
  \be\left\{\ba{lcl}
  \xi &\lra& 0,\\
  \tilde{H} & & invariant.
  \ea\right .\ee
This gives us 2 extra Jacobians:\\
 the first
\be
J_2 = \prod_x\ (0),
\ee
comes from the transformation of ${\cal D}\xi$, but goes away in
the overall normalization of $Z$; the fact that the
transformation on $\xi$ is singular (in (\cite{ab-lee})
 as well as here) should
consequently not appear as a problem in the computation of Green
functions;\\
the second comes from the transformation of the fermionic measure
\cite{fuji}
\be
J = e^{i(\xi/v){\cal A}},
\ee
where $\cal A$ is the anomaly.\\
We can now rewrite Z as
\be
Z = \int {\cal D} \Psi{ \cal D} \overline\Psi
{\cal D} \tilde H {\cal D} \xi {\cal D} \sigma_\mu
\  J_1 J_2\   e^{i\int d^4 x \ti{\cal L}(x)+ {\cal L}_c(x)
+(\xi/v){\cal A}}, \ee
with
\be\ba{ccl}
\ti{\cal L}& =&
 -{1\over 4}F_{\mu\nu}F^{\mu\nu} +i\overline\Psi\gamma^\mu
 \left( \p_\mu -ig\sigma_\mu {\Bbb T}_L\right)\Psi
 +(\xi/v){\cal A}-(\xi/v)\p_\mu J^\mu_\psi\\
 & &
 +{1\over 2}\p_\mu\tilde H\p^\mu\tilde H +{1\over
 2}\sigma_\mu^2 \tilde H^2 -V(\tilde H^2).
 \ea\la{eq:Ltilde}\ee
As $J_1$ and $J_2$ do not depend on $\xi$, the $\xi$ equation
is now
\be
\p_\mu J^\mu_\psi -{\cal A} - v{\p{\cal L}_c\over \p\xi}=0.
\la{eq:xieq}\ee
$v\ {\p{\cal L}_c /\p\xi}$ is precisely the contribution $\p_\mu
J^\mu_{\psi c}$ to the divergence of the fermionic current
coming from ${\cal L}_c$, as can be seen when varying the
Lagrangian with a global fermionic transformation, such that
(\re{eq:xieq}) rewrites
\be
\p_\mu J^\mu_\psi -{\cal A} -\p_\mu J^\mu_{\psi c}=0,
\ee
and is nothing more than the exact (quantum) equation for
$\p_\mu J^\mu_\psi$. (We shall furthermore show below that the
fermionic current is exactly conserved: $\p_\mu J^\mu_{\psi c}$
involves a classical contribution
$(\p_\mu J^\mu_\psi)_{classical}$ which vanishes plus a
quantum contribution exactly cancelling the anomaly.)\\
In this form of $Z$, $\xi$ appears as a Lagrange multiplier;
the associated constraint being satisfied at all orders, $\xi$
disappears: it has been "gauged away" and transformed into the third
polarization of the massive vector boson.\\
This form of the theory describing a massive gauge field is
manifestly unitary. The gauge invariance (\re{eq:qsym}) has
vanished; the transformation (\re{eq:chvar}) is equivalent to
going to the unitary gauge.\\
Wether or not all 3 polarizations of the massive vector field should be
considered as composite is certainly a legitimate question; however, we
do not intend to consider this problem here.\\
As expected, this form of the theory is not suitable for
looking at renormalizability.

  \subsection{Renormalizability.}

We work with the theory defined with ${\cal L}+{\cal L}_c+{\cal
L}_{gf}+{\cal L}_{FP}$ in terms of the variables $H$ and
$\phi$.\\
The 2 potential sources of non-renormalizability are:\\
\quad - the 4-fermi couplings in ${\cal L}_c$;\\
\quad - the coupling $(\xi/v)\ \p_\mu J^\mu_\psi$.\\
We will show in the following that the infinite mass given by
${\cal L}_c$ to the fermions when $H$ gets its vacuum expectation
value $v$ has drastic consequences on both:\\
- the fermionic current gets conserved and thus its derivative coupling
to $\xi$ harmless;\\
- working at the one-loop level, infinite series of "ladder" diagrams
can be resummed, leading to effective renormalizable
4-fermions couplings.\\
The infinite quark mass
\be
m_0 = \lim_{\beta\ra 0}\ - \frac{\Lambda^2 v^2}{\beta\mu^3}.
\la{eq:m0}\ee
is, as we show below, stable by the above resummation.\\
Its first non trivial effect is to cancel the anomaly in
$\p_\mu J^\mu_\psi$. It can be most easily seen in the
Pauli-Villars regularization of the triangular diagram (see
fig.(4)), since the Ward Identity reads \cite{ano}:
\be
k^\mu\ (T_{\mu\nu\rho}(m_q) -T_{\mu\nu\rho}(M))= m_q\ T_{\nu\rho}
(m_q) - M\ T_{\nu\rho}(M),
\ee
where $m_q$ is the effective quark mass (that we shall see
below to be the same as $m_0$) and $M$ is the mass of
the regulator.\\
As the anomaly comes from
\be
{\cal A}=\lim_{M\ra \infty}\ -M\ T_{\nu\rho}(M),
\ee
it is now exactly cancelled by $m_q\ T_{\nu\rho}(m_q)$ as $m_q$
goes to infinity when $\beta$ goes to $0$.\\
We conclude that quantum corrections to $\p_\mu J^\mu_\psi$
cancel and that this operator is identical to its classical
expression
\be
(\p_\mu J^\mu_\psi)_{classical}=
-i{m_q\over v}(H\ol\Psi\gamma_5{\Bbb T}\Psi -\phi\ol\Psi\Psi)
_{classical}.
\ee
The above expression identically vanishes by the classical
equations for $H$ and $\phi$ at the limit $\beta\ra 0$.\\
We just showed that both the classical and the quantum
corrections to the divergence of the fermionic current get
cancelled, and so that this current is conserved. Its
problematic coupling to $\xi$ becomes consequently harmless.
\newpage\null

\vskip 6truecm
{\centerline{\em Fig.4: Triangular diagrams involved in the
anomalous Ward Identity.}}
\vskip 5mm
The second step is to prove the renormalizability of the
4-fermi couplings in ${\cal L}_c$. We shall only  make the demonstration
 at the one-loop level.\\
Their bare values are infinite when $\beta\ra 0$:
\be
\zeta_0 = -\zeta_0^5 =
-\frac{\Lambda^2v^2}{2\beta\mu^6}={m_q\over 2\mu^3},
\ee
but are modified by the resummation of the infinite series of
1-loop 1PI diagrams depicted in fig.(5), accordind to
\be
\zeta_0 \lra \zeta(q^2)=\frac{\zeta_0}{1-\zeta_0\  A(q^2,m_q)}.
\la{eq:zeta}\ee
 \vskip 5truecm
 {\centerline{\em Fig.5: The effective 4-fermions coupling
 $\zeta(q^2)$.}}
 \vskip 5mm
\noindent $A(q^2,m_q)$ is the 1-loop fermionic bubble,
 and $m_q$ is the
effective fermion mass obtained by the resummation depicted in
fig.(6):
\be
m_q= m_0 -2\zeta(0)\ \mu^3.
\la{eq:mq}\ee
\newpage\null
\vskip 5truecm
{\centerline{\em Fig.6: Resumming the fermion
propagator.}}
\vskip 5mm
\noindent (Note that the "non-leading" contributions
 to $m_q$ coming from \linebreak
 $(\overline\Psi\Psi)^2$ and $(\overline\Psi\gamma_5{\Bbb
T}\Psi)^2$ cancel -see fig.(7)-).
%\newpage
%\null
\vskip 5truecm
{\centerline{\em Fig.7: Cancellation of the "non-leading"
contributions to the fermion mass.}}
\vskip 5mm

\noindent From (\re{eq:zeta}) we get $\zeta(0)$ and, so, from
 (\re{eq:mq})
\be
m_q= m_0 - \frac{2\zeta_0\ \mu^3}{1-A(0,m_q)\ \zeta_0}.
\la{eq:meq1}\ee
Once renormalised, $A(q^2,m)$ behaves for high $q^2$ like (see
\cite{broa} for example)
\be
A(q^2,m) = \alpha q^2 + \delta m^2 +\cdots,
\ee
yielding for $m_q$ the equation
\be
(m_q-m_0)(1-\delta m_q^2\ \zeta_0)= -2\zeta_0\ \mu^3=-m_0.
\la{eq:meq2}\ee
When $\beta\ra 0$, (\re{eq:meq2}) has the solutions
\be
\left\{\ba{lcl}
  m_q &= & 0,\\
  m_q &= & m_0\ +{\cal O}(\beta^2),\\
  m_q &= & 0\ -{\cal O}(\beta^2).
  \ea\right .
  \la{eq:msol}\ee
We discard the solutions $m_q=0$ which leave the anomaly
uncancelled: the fermionic current would stay not conserved
and the theory non-renormalizable because of the  $(\xi/v)\
\p_\mu J^\mu_\psi$ coupling.\\
The stable solution $m_q=m_0\ +{\cal O}(\beta^2)$ corresponds to
\be
\lim _{\beta\ra 0}^{q^2\ra \infty} \zeta(q^2) =-{1\over{\alpha
q^2 +\delta m_q^2}}.
\ee
We thus conclude that, accordingly:\\
\quad - the effective coupling $\zeta(q^2)$ behaves like
$1/q^2$ when $q^2$ becomes large;\\
\quad - it furthermore vanishes like $\beta^2$.

We can show in the same way that the effective $\zeta^5(q^2)$
behaves like\linebreak
 $-1/(\alpha_5 q^2 +\delta_5 m_q^2 +B_0)$, where
$B_0$  is the nonperturbative contribution to the fermionic
bubble \cite{broa}.
\medskip

This shows that the perturbative series built with the
effective 4-fermi couplings $\zeta(q^2)$ and $\zeta^5(q^2)$ has
all the right properties for renormalizability; the effective
4-fermi couplings get dimension (-2) and the
 counterterms, not initially present in the Lagrangian,
 that could be expected in the renormalization of the one loop diagrams
depicted in fig.(8), corresponding to 6-fermions,  8-fermions,
4 or 6-fermions-gauge field or 4 or 6-fermions-scalar interactions
all go to $0$ like powers of $\beta$.\\
\vskip 8 truecm
{\centerline{\em Fig.8: Possible contributions to counterterms vanishing
when $\beta\ra 0$.}}
\vskip 5mm

It is clear that if all original diagrams are present in the reshuffled
perturbative series built with $\zeta(q^2)$ and $\zeta^5(q^2)$, extra
diagrams appear, of the type depicted in fig.(9). However the $\beta^2$
dependance of the effective couplings makes them also
vanish like powers of $\beta$.
\newpage\null
\vskip 5truecm
{\centerline{\em Fig.9: Extra diagram generated by expanding in powers
of $\zeta(q^2)$ and $\zeta ^5(q^2)$.}}
\vskip 5mm

We conclude this section by stressing again the role of the
constraints for the renormalizability of the theory though the
infinite mass given to the fermions. We hope to be able to present
a demonstration at all orders in the next future.

We turn now to more phenomenological consequences.

\section{Particles.}

We have already seen that an effect of the constraint is to
give  the fermions (quarks) constituting the scalars an
infinite mass . They are consequently not expected to be
produced as asymptotic states, and only to propagate in internal
lines, producing renormalization effects \cite{ac}. (One of the
most spectacular is, as shown above, the cancellation of the
anomaly).
Its other dramatic effect concerns the Higgs particle.

\subsection{An infinite mass for the Higgs field.}

The scalar potential, initially choosen as
\be
V(H,\phi)= -{\sigma^2\over 2}(H^2-\phi^2)+{\lambda\over
4}(H^2-\phi^2)^2
\ee
is itself modified by the constraint in its exponentiated form
to become
\be\ba{lcl}
\ti V(H,\phi)&=& V(H,\phi)+\lim_{\beta\ra 0}\\
& &{\Lambda^2\over 2\beta}\left(H^2 -\phi^2 -{2v\over
\mu^3}(H\overline\Psi\Psi-\phi
\overline\Psi\gamma_5 {\Bbb T} \Psi)+
 {v^2\over\mu^6}\left((\overline\Psi\Psi)^2
-(\overline\Psi\gamma_5 {\Bbb T} \Psi)^2\right)\right).
\ea\la{eq:Vtilde}\ee
Its minimum still corresponds to $\langle H\rangle = v$,
$\langle \overline\Psi \Psi \rangle =\mu^3$, $\langle \phi\rangle
= 0$ if $\sigma^2 =\lambda v^2$, but the Higgs mass has now
become
\be
\left .\frac{\p^2\ti V}{\p H^2}\right|_{H=v}= -\sigma^2 + 3\lambda
v^2 +{\Lambda^2\over\beta},
\ee
which goes to $\infty$ at the limit $\beta\ra 0$ when the
constraints are implemented.\\
We conclude that the Higgs field cannot be produced either as
an asymptotic state.

\subsection{Counting the degrees of freedom.}

It is instructive at that point to see by which mechanism the degrees
of freedom have been so drastically reduced, since neither the Higgs
nor the quarks are expected to be produced.\\
We have started with 2 (2 scalars) + 4 (one vector field) + 4N (N
fermions) degrees of freedom. They have been reduced to only 3 (the 3
polarizations of a massive vector boson, the 3rd of which being the
pion as we shall see below) by 4N + 3 constraints which are the following:\\
- the 2 constraints linking $\phi$ and $H$ to the fermions;\\
- the gauge fixing allowed by gauge invariance;\\
- the 4N constraints coming from the condition
$\langle\overline\Psi\Psi\rangle =\mu^3$: indeed because of the
underlying fermionic $O(N)$ invariance of the theory
and Lorentz invariance, this condition is equivalent to
\be
 \langle\overline\Psi ^\alpha_n\Psi^\alpha_n\rangle =0,
\  for\  n=1 \ldots\  N\  and \ \alpha\ =1\ \ldots\  4,
\ee
which make 4N equations.\\
 The latter are specially important since
 the fermions become infinitely massive precisely because of
the condition on $\langle\overline\Psi\Psi\rangle$  as can be seen by
looking at (\re{eq:Lc}).\\
In our scheme, the breaking of the gauge symmetry is also necessary for
the fermions making up the scalars not to appear as asymptotic states.

\subsection{The scalar $\phi$ as an abelian pion; introducing
leptons.}

We introduce leptons in our theory by adding to $\cal L$, by
analogy with the quarks:
\be
 i\overline\Psi_\ell\gamma^\mu\left(\p_\mu -ig(\sigma_\mu
-{1\over g}{\p_\mu\xi\over v}){\Bbb T}_L\right)\Psi_\ell,
 \la{eq:lept}\ee
 where $\Psi_\ell$ is now an $N_\ell$ vector and ${\Bbb
 T}_L$ an
 $N_\ell\times N_\ell$ matrix ($N_\ell$ can be different
from $N$). This involves in
particular the coupling of $\xi$ to leptons
\be
{\xi\over v}\ \p^\mu J_{\mu\ell}^\psi.
\ee
 where
\be J^\psi_{\mu\ell} =\overline\Psi_\ell \gamma_\mu {\Bbb T}_L
 \Psi_\ell ={1\over 2}(V_{\mu\ell}-J^5_{\mu\ell}).
 \ee
We define as usual the constant $f_\pi$ controlling the
leptonic pion decays by
\be
\langle 0\mid J^\mu_5(0)\mid\pi(k)\rangle_{in} = i\sqrt{2} f_\pi k^\mu,
\ee
leading to
\be
\langle 0\mid\p_\mu J^\mu_5\mid\pi\rangle = -\sqrt{2} k^2 f_\pi.
\la{eq:pcac}\ee

The equation for $\phi$ coming from ${\cal L}+{\cal L}_c$ is
\be\ba{lcl}
D^2\phi&=& -{1\over v}{\p\xi\over\p\phi}\p_\mu(J^\mu_\psi
+J^\mu_{\psi\ell})-{\p{\cal L}_c\over\p\phi}\\
       &=& {i\over 2v}(1+{h\over v}+\cdots)\p_\mu(J^\mu_5
+J^\mu_{5\ell}) -{\Lambda^2\over \beta}(\phi-{v\over\mu^3}\ol\Psi
\gamma_5{\Bbb T}\Psi),
\ea\la{eq:eqphi}\ee
where we have used the fact that the vector currents are
conserved.\\
Plugging (\re{eq:eqphi}) into (\re{eq:pcac}) and using
\be
\langle 0\mid \p_\mu J^\mu_{5\ell}\mid\pi\rangle =0,
\ee
we get
\be
\langle 0\mid
-2iv(D^2\phi(1-{h\over v}+\cdots)+{\Lambda^2\over\beta}
(\phi-{v\over\mu^3}\ol\Psi\gamma_5{\Bbb T}\Psi))\mid\pi\rangle=
-\sqrt{2}k^2f_\pi.
\ee
The $1/\beta$ term only implements the constraint
$\phi={v\over\mu^3}\ol\Psi\gamma_5{\Bbb T}\Psi$
and we get the relation
\be
\langle 0\mid D^2\phi(1-{h\over v}+\cdots)\mid\pi\rangle=
 -{i\over v\sqrt{2}}f_\pi k^2,
\ee
or
\be
\langle 0\mid (k^2 \phi -4ig\sigma^\mu\p_\mu H -2igh \p^\mu\sigma_\mu
+2g^2\sigma_\mu^2 \phi)(1-{h\over v}+\cdots)
\mid\pi\rangle ={i\over v\sqrt{2}}f_\pi k^2,
\ee
yielding in particular
\be
\langle 0\mid\phi(1-{h\over v}+\cdots)\mid\pi\rangle=
{i\over v\sqrt{2}} f_\pi.
\ee
As we have normalized $\phi$ to 1 (and not $\pi$), this
leads to
\be
\phi(1-{h\over v}+\cdots) =-i \sqrt{2} {v\over f_\pi} \pi.
\la{eq:phipi}\ee
Using (\re{eq:phipi}) we have
\be\ba{lcl}
_{out}\langle\ell\overline\ell\mid\pi\rangle_{in}&=&
i{f_\pi\over\sqrt{2}v}\
{_{out}\langle\ell\overline\ell\mid\phi\rangle_{in}}\\
&=&
i{f_\pi\over\sqrt{2}v}\
\langle\ell\overline\ell\mid Te^{iS}\mid\phi\rangle\\
&=&
{g^2\over\sqrt{2}M^2}f_\pi k_\mu\
\langle\ell\overline\ell\mid J^\mu_\ell\rangle.
\ea\la{eq:pilept}\ee
In the last line of (\re{eq:pilept})
we have used (\re{eq:gmass}).
We recover the same result as given by the usual PCAC
computation. $\phi$ is seen to behave like an abelian pion.

It is also worth mentionning that, because of the cancellation
of the anomaly in the quark sector (see above), the decays of
$\phi$ into 2 "photons" now occur through leptonic triangular
diagrams. The leptonic current remains indeed anomalous.

Because in particular of this leptonic anomaly, the leptonic
sector as described above stays non-renormalizable. This will be
dealt with in a forthcoming work \cite{bm1}.

\subsection{Summary: asymptotic states.}

We have already seen that the quarks and the Higgs field get
infinite masses from the constraints, and thus cannot appear as
asymptotic states.
The only true "particles" left are consequently the massive
gauge field $\sigma_\mu$, the third polarization of which behaves
exactly like a (abelian) pion, and the leptons.\\

\section{Conclusion and perspectives.}

 We have discussed at length above a way of dealing with a
certain type of
 spontaneously broken $U(1)_L$  gauge theories.  The
first remark to be made is that we do
not know what to do when the symmetry is not spontaneously
broken, as everything rests on the existence of a scale $v$
characterizing the vacuum expectation value of a
scalar field. When $v\ra 0$, the change of variables
(\re{eq:chvar1})  becomes singular, meaning that the fields
$\tilde H$ and $\xi$ cannot be defined;
the compensating field with
dimension 1 has been intimately linked with the presence
of a massive gauge field. \\
A second point that we want to stress is that we have
given up the gauge invariance of the action itself; we have instead
made its non-invariance cancel that of the fermionic
measure in the generating functionnal. It can be
argued that our results may not be true at the operator level,
but only at the level of vacuum expectation values (or Green
functions). \\
The phenomenological consequences of our approach are
nontrivial since the quarks and the Higgs are now unobservable.\\
We also showed that the pseudoscalar partner $\phi$ of the Higgs
can be "gauged into" the 3rd polarization of the massive vector boson,
and that it can be identified with an abelian pion;
in particular, no other extra scale of interaction needs to be
introduced, unlike in "technicolour" theories.\\
The generalization to the non-abelian case in under
investigation \cite{bm2}, and uses in particular the presence,
in the Standard Model, of an associative algebra which allows
the same construction as above.\\
While the quark sector has been extensively studied, we have
left the leptonic sector plagued with anomalies.
 We shall show in a forthcoming work how
this problem could be circumvented \cite{bm1}.\\

\vskip 1.5cm

{\em{\underline {Aknowledgements}}:
it is a pleasure to thank my colleagues at LPTHE in
Paris, and the warm hospitality that I always received at the
Institut f\"ur Physik in Mainz. Special thanks are due to
George Thompson for his numerous advice and critics
(though the author takes the full responsability for the errors
that can be present in the paper), and to Professor Karl Schilcher.
\newpage
\null

\appendix
{\Large\bf Appendix A: The gauge invariance (\re{eq:qsym})}
\bigskip

Let us work in the form of the theory where the constraints are not
exponentiated. The action then corresponds to massless fermions and,
consequently, the only contribution to the divergence of the fermionic
current is the anomaly, according to
\be \partial^\mu J_\mu^\psi ={\cal A}.\la{eq:djano}\ee
This is the direct consequence of expressing the invariance of
 $Z$, defined by the eqs.(\re{eq:Z}) and (\re{eq:L}), by the
transformation (\re{eq:qsym})  because the action $S$ gets transformed into
\be S - \int d^4x\  \theta\   \partial^\mu J_\mu^\psi ,\ee
and the fermionic measure into
\be {\cal D}\Psi {\cal D}\overline\Psi\  e^{\int d^4x\  \theta {\cal
A}}.\ee

To study the Green functions of the theory, we must
introduce sources for all the fields, adding to the Lagrangian
\be \overline J\Psi + \overline \Psi J + K\xi +L\tilde H -g
R_\mu\sigma^\mu,\la{eq:source}\ee
and look at the Ward Identities resulting from the invariance of
$Z(sources)$ by
(\re{eq:qsym}).  The vanishing of the variation of $Z(sources)$ yields
\be \langle -\partial^\mu J_\mu^\psi +{\cal A} -Kv -\partial^\mu R_\mu
-i\overline J {\Bbb T} {1-\gamma_5\over 2}\Psi +i\overline\Psi
{1+\gamma_5\over 2}{\Bbb T} J \rangle = 0 .\la{eq:varnul}\ee
As all the sources are finally to be put to 0, the $Kv$ term will play
no role as far as the Ward Identities are concerned.
Using (\re{eq:djano}), we recover from (\re{eq:varnul}) the Ward
Identities that would occur in an anomaly free theory, as can be easily
seen by making a transformation on $\Psi$, $\overline\Psi$ and
$\sigma_\mu$ only in a "standard" theory with no $\xi$ field.
 Those naive Ward Identites deduced from gauge invariance
are usually known to be broken by the anomaly; we see here that they
are still valid in the presence of the anomaly, thanks to the
introduction of the $\xi$ field. By (\re{eq:djano}) the non-invariance
of the action $\langle -\partial^\mu J_\mu^\psi\rangle$
 has been compensated by that of the fermionic measure
$\langle{\cal A}\rangle$.

This is precisely what is meant by the invariance (\re{eq:qsym}).

Let us  notice that
\be \langle Kv +\partial^\mu R_\mu \rangle = 0 \ee
is the identity which would result from the invariance (\re{eq:clasym})
if it were satisfied.

\newpage
\null

\appendix
{\Large\bf Appendix B:  The Faddeev-Popov procedure}
{\rm\cite{fadpop}}
\vskip 1cm
\noindent Let the theory be defined by
\be
Z=\int{{\cal D}\Psi{\cal D}\ol\Psi{\cal D}H{\cal D}\phi
{\cal D}\sigma_\mu
\ e^{i\int d^4x {\cal L}(\Psi,H,\phi,\sigma)}
\ \delta(C_H)\ \delta(C_\phi)}.
\la{eq:Zap}\ee
We introduce as usual $1$ into $Z$ in the form
\be
1=\Delta(\sigma)\int{{\cal D}{\cal G}\  \delta(F(\sigma^g))},
\ee
where $\sigma$ is the gauge field and $\sigma^g$ its
transformed by the gauge transformation (\re{eq:qsym});
the integration is made over the gauge group $\cal G$.\\
Writing $H^g$, $\phi^g$, $g^{-1}\Psi$ the gauge transformed of
$H$, $\phi$ and $\Psi$ by (\re{eq:qsym}), (\re{eq:Zap}) can be
rewritten
\be\ba{lcl}
 Z&=&\int{{\cal D}\Psi{\cal D}\ol\Psi{\cal D}H{\cal D}\phi{\cal
 D}\sigma_\mu{\cal D}{\cal G}}\\
& &\ e^{i\int d^4x {\cal L}(g^{-1}\Psi,H^g,\phi^g,\sigma^g)
 -\theta\,\p_\mu J^\mu_\psi}\
\delta(C_H)\ \delta(C_\phi)
 \ \Delta(\sigma) \ \delta(F(\sigma^g)).
\ea\ee
We use now
\be
\Delta(\sigma)=\Delta(\sigma^g),\quad
{\cal D}\sigma={\cal D}\sigma^g,\quad
{\cal D}H^g\  {\cal D}\phi^g={\cal D}H\  {\cal D}\phi,
\ee
the transformation of the fermionic measure \cite{fuji}
\be
{\cal D}\Psi{\cal D}\ol\Psi=
{\cal D}(g^{-1}\Psi){\cal D}(g^{-1}\ol\Psi)
\ e^{i\int {\theta{\cal A}}},
\ee
and the invariance of the product $\delta(C_H)\  \delta(C_\phi)$
by a transformation acting on both the fermions and the scalars.
Going to the new variables $\sigma^g, H^g, \phi^g, g^{-1}\Psi$
and using the equation (\re{eq:djano}) deduded in Appendix A, (\re{eq:Zap})
rewrites
\be
 Z=\int{{\cal D}\Psi{\cal D}\ol\Psi{\cal D}H{\cal D}\phi{\cal
 D}\sigma_\mu{\cal D}{\cal G}
\ e^{i\int d^4x {\cal L}(\Psi,H,\phi,\sigma)}
\ \delta(C_H)\ \delta(C_\phi)\ \Delta(\sigma) \ \delta(F(\sigma))}.
\ee
This shows in particular that the integration over the gauge
group factorizes.

\newpage\null
\listoffigures\noindent
 Fig.1: Cancellation of the ghost residue by the Goldstone in
 the Landau gauge;\\
 Fig.2: Resumming the $\sigma_\mu$ and $\xi$ propagators;\\
 Fig.3: Other diagrams for ghost cancellation in a general
 $\alpha$-gauge;\\
 Fig.4: Triangular diagrams involved in the anomalous Ward
Identity;\\
 Fig.5: The effective 4-fermion coupling $\zeta(q^2)$;\\
 Fig.6: Resumming the fermion propagator;\\
 Fig.7: Cancellation of the "non-leading" contributions to the
fermion mass;\\
 Fig.8: Possible contributions to counterterms vanishing
when $\beta\ra 0$.\\
 Fig.9: Extra diagram generated by expanding in powers
of $\zeta(q^2)$ and $\zeta ^5(q^2)$.

\newpage\null

\end{document}